\documentclass[twocolumn,showpacs,
amsmath,amssymb]{revtex4}
\usepackage{amssymb}
\usepackage[dvips]{graphicx}
\begin{document}

\title{Unconventional superconducting pairing by conventional phonons}
\author{A. S. Alexandrov}

\affiliation{Department of Physics, Loughborough University,
Loughborough LE11 3TU, United Kingdom\\}

\begin{abstract}
The common  wisdom that the  phonon mechanism of electron pairing in
the weak-coupling Bardeen-Cooper-Schrieffer (BCS) superconductors
leads to conventional s-wave Cooper pairs is revised. An inevitable
anisotropy of  sound velocity  in crystals makes the phonon-mediated
attraction of electrons non-local in space providing unconventional
Cooper pairs with a nonzero orbital momentum in a wide range of
electron densities. As a result of this anisotropy quasi-two
dimensional charge carriers undergo a quantum phase transition from
an unconventional d-wave superconducting state to a conventional
s-wave superconductor with more carriers  per unit cell. In the
opposite strong-coupling  regime rotational symmetry breaking
appears as a result of a reduced Coulomb repulsion between
unconventional bipolarons dismissing thereby some constraints on
unconventional pairing in the Bose-Einstein condensation (BEC)
limit. The conventional phonons, and not superexchange, are shown to
be responsible for the d-wave symmetry of  cuprate superconductors,
where the on-site Coulomb repulsion  is large.

\end{abstract}


\maketitle

A well-known quantum mechanics theorem  \cite{landau} states that
the coordinate wave function of  two particles does not become zero
(or has no nodes) in the ground state. If it has no nodes  the wave
function must certainly be symmetrical with respect to the
interchange of the particle coordinates, ${\bf r}_1 \rightleftarrows
{\bf r}_2$; for, if it were to be antisymmetrical, it would vanish
at ${\bf r}_1= {\bf r}_2$. Thus the system of two identical fermions
must have a spin-singlet ground state, as the complete wave function
(which includes  spin coordinates) must be antisymmetrical.

 It has been thought for a
long while that  Cooper pairs in the BCS theory with the
conventional
 electron-(acoustic) phonon interaction (EPI) are singlets and
their wave function is isotropic (s-wave), in apparent agreement
with the theorem \cite{bcs}. The interaction has been thought to be
local
 in space, so it could not
 lead to a higher angular momentum pairing. If the pairing
interaction is non-local,   the BCS theory admits  unconventional
Cooper pairs with nonzero orbital momentum  \cite{lif}, as the
theorem is not
 applied to  more than two particles.  Thus it has gone unquestioned that the unconventional pairing
 requires unconventional electron-phonon
 interactions  with specific  phonons and  poor screening
 \cite{abr,sig,shen,kulic,tsai},
  sometimes combined with anti-ferromagnetic fluctuations \cite{mazin} and  vertex corrections \cite{hag},
 or  non-phononic
 mechanisms of pairing (e.g. superexchange \cite{band}), and a specific shape of the Fermi surface. Since the rotational symmetry breaking is
a many-body effect, any superconductor  should seem to be s-wave
 in the strong-coupling  limit
\cite{bristol}, where  pairs are  individual (e.g. bipolarons
\cite{alebook})  rather than overlapping  Cooper pairs.

Here the symmetry of the superconducting  state mediated by
 conventional acoustic phonons is reconsidered. The sound speed anisotropy leads to a
double surprise: a) the BCS  state of  layered crystals is d-wave in
a wide range of carrier densities, b) the strong-coupling BEC
 state  can  break the rotational symmetry as well as the
weak-coupling BCS state.

In the framework of the BCS theory the symmetry of the order
parameter $\Delta(\bf k)$ and the critical temperature, $T_c$, are
found by solving the linearised "master" equation \cite{bcs},
\begin{equation}
\Delta({\bf k})=-\sum_{\bf k^{\prime }}V({\bf k,k^{\prime}}){\frac{\Delta ({\bf k^{\prime} })}{2\xi _{{\bf %
k^{\prime }}}}}\tanh \left({\xi _{\bf %
k^{\prime }}\over{2k_BT_c}}\right). \label{master}
\end{equation}
The interaction $V(\bf k,k^{\prime })$ comprises the attraction, $-
V_{ph}({\bf q})$, mediated by acoustic phonons, and the Coulomb
repulsion, $V_{c}({\bf q})$, as
\begin{eqnarray}
V({\bf k,k^{\prime}})&=&-V_{ph}({\bf q})\Theta(\omega_D-|\xi _{\bf
k}|)\Theta(\omega_D-|\xi _{\bf k^{\prime}}|)\cr &+&V_{c}({\bf
q})\Theta(\omega_p-|\xi _{\bf k}|)\Theta(\omega_p-|\xi _{\bf
k^{\prime}}|),
\end{eqnarray}
where $V_{ph}({\bf q})=C^2/NMc_l^2$ is  the square of the matrix
element of the electron-phonon interaction \cite{bardeen} divided by
the square of the acoustic phonon frequency, $\omega_{\bf q}=c_l q$,
$c_l$ is  sound velocity, $M$ is the ion mass,  $N$ is  the number
of unit cells in the crystal, and $\xi _{\bf k}$ is the electron
energy relative to  the Fermi energy. The deformation potential
matrix element  $C$ is roughly the electron bandwidth in  rigid
metallic \cite{bardeen}, or semiconducting  \cite{anselm} lattices,
${\bf q}={\bf k -k^{\prime}}$ is the electron momentum transfer with
the magnitude $q= 2^{1/2}k_{F}[1-\cos \Phi ]^{1/2}$, where $\Phi $
is the angle between ${\bf k}$ and ${\bf k^{\prime }}$, and $\hbar
k_F$ is the Fermi momentum. Theta functions in Eq.(\ref{master})
($\Theta(x)=1$ for positive $x$ and zero otherwise) account for a
difference in frequency scales of the electron-phonon interaction,
$\omega_D$, and the Coulomb repulsion, $\omega_p\gg \omega_D$, where
$\omega_D$ and $\omega_p$ are the maximum  phonon and plasmon
energies, respectively.

If one neglects anisotropic effects, replacing $V_{ph}({\bf q})$ and
$V_{c}({\bf q})$ by their Fermi-surface averages, $V_{ph}({\bf
q})\Rightarrow V_{ph}$, $V_{c}({\bf q})\Rightarrow V_c$ \cite{bcs},
then there is only an $s$-wave  solution of Eq.(\ref{master}),
$\Delta_s$, independent of ${\bf k}$.  The sound speed anisotropy
actually changes the symmetry of the BCS state. While $c_l$ is a
constant in the isotropic medium, it depends on the direction of
${\bf q}$  in any crystal. The anisotropy is particulary large in
layered crystals like cuprate superconductors, where an elastic
stiffness constant in the $a-b$ plane is substantially greater than
in the $c$ direction (see \cite{migliori,chang} and references
therein). As an example,  the measured velocity of longitudinal
ultrasonic waves along $a-b$ plane, $c_{\parallel}$=4370 ms$^{-1}$
 is almost twice larger than that along $c$ axis, $c_{\perp}$=2670 ms$^{-1}$
in Bi$_2$Sr$_2$CaCu$_2$O$_{8+y}$ \cite{chang}. It makes $V_{ph}({\bf
q})$ anisotropic,
\begin{equation}
V_{ph}({\bf q})={C^2\over{NMc_{\perp}^2(1+\alpha
q_{\parallel}^2/q^2)}}, \label{ph}
\end{equation}
where $\alpha=(c_{\parallel}^2-c_{\perp}^2)/c_{\perp}^2$ is the
anisotropy coefficient, which is about $2$ in  cuprates. Also the
Coulomb repulsion is $q$ dependent, $ V_c({\bf q})=4\pi
e^2/V\epsilon_0(q^2+q_s^2). $ Here $\epsilon_0$ is the dielectric
constant of the host lattice of the volume $V$, and $q_s^2=8\pi
e^2N(0)/V\epsilon_0$ is the inverse screening radius squared with
the density of states (per spin), $N(0)$, at the Fermi surface.

Now we can solve the master equation (\ref{master}) by   expanding
the order parameter and the potentials in   series of the spherical
harmonics. If the electron energy spectrum is quasi-two dimensional
(2D) as in  cuprates, $\xi _{\bf k}\approx \xi _{\bf
k_{\parallel}}$, one can   expand $\Delta({\bf k})=\sum_m \Delta_m
\exp(im\phi)$ and $V_{ph,c}({\bf q})=\sum_{m}V_{ph,c}(
q_{\perp},m)\exp[im(\phi-\phi^{\prime})]$ in series of the
eigenfunctions of the $c$-axis component of the orbital angular
momentum, where $\phi$ and $\phi^{\prime}$ are the polar angles of
${\bf k}_{\parallel}$ and ${\bf k}^{\prime}_{\parallel}$,
respectively.

\begin{figure}
\begin{center}
\includegraphics[angle=-90,width=0.50\textwidth]{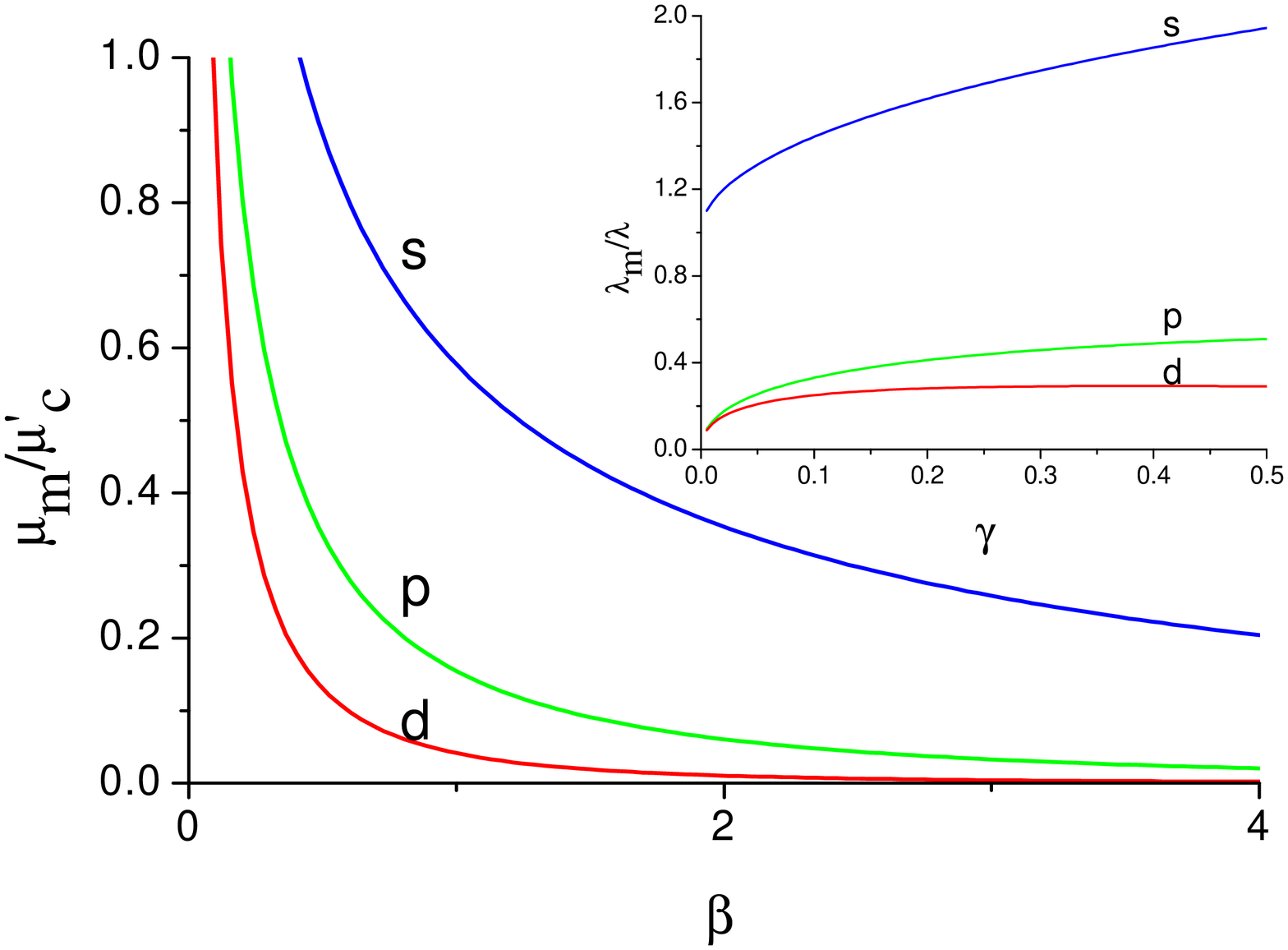}
\vskip -0.5mm \caption{The Coulomb repulsion, $\mu_{m}$, as a
function of the ratio of the electron wavelength to the screening
length squared ($\beta=q_S^2/2k_F^2$), and the electron-phonon
coupling constant, $\lambda_m$,  as a function of the ratio of the
electron wavelength to the inter-plane distance squared,
$\gamma=\pi^2/2d^2k_F^2(1+\alpha)$ for $\alpha=4$ (inset) in $s,p$
and $d$ pairing channels. Here $\mu_c^{\prime}=\mu_c
\tilde{\gamma}$. }
\end{center}
\end{figure}

\begin{figure}
\begin{center}
\includegraphics[angle=-90,width=0.50\textwidth]{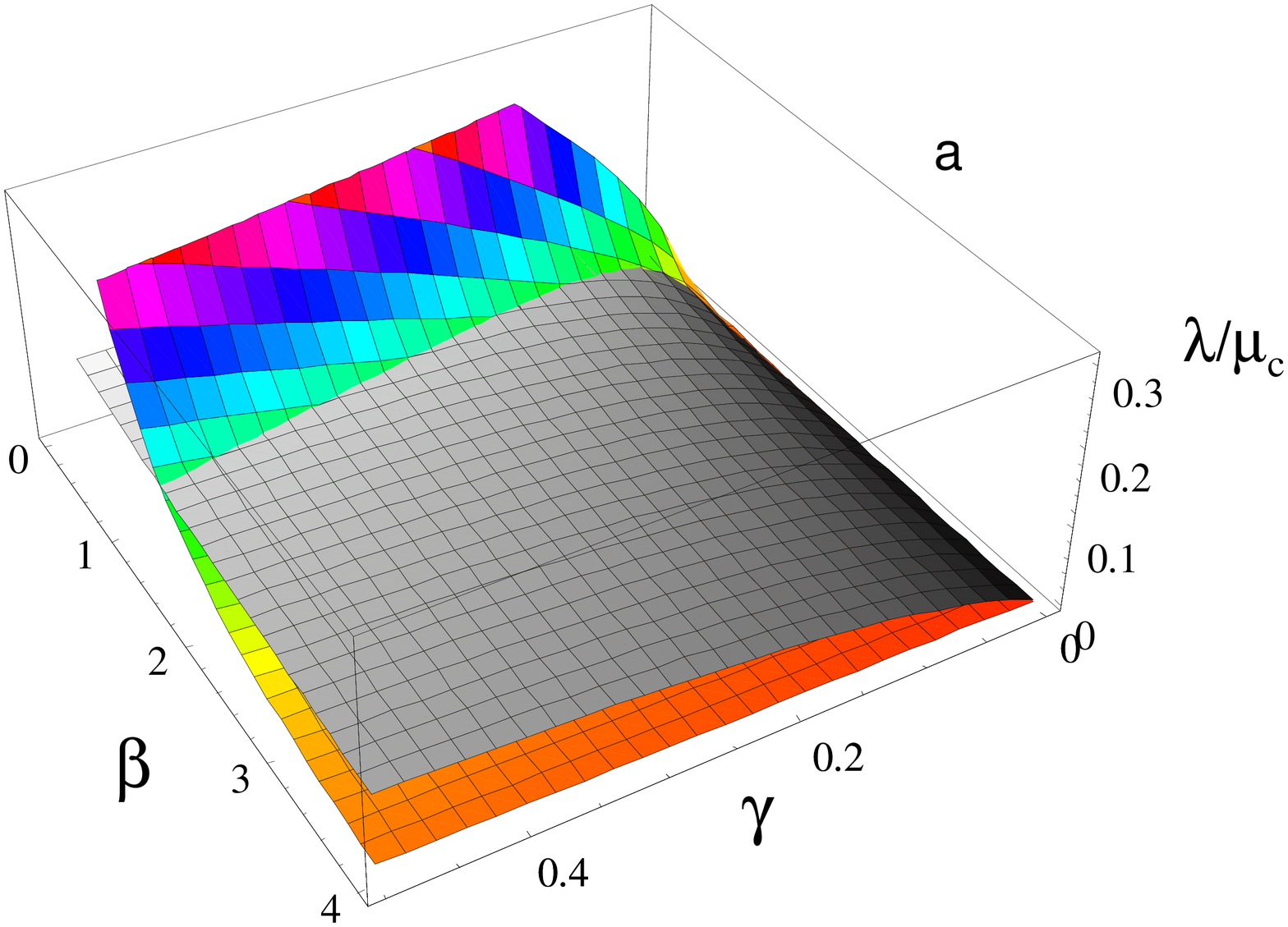}\\[0.2cm]
\includegraphics[angle=-90,width=0.50\textwidth]{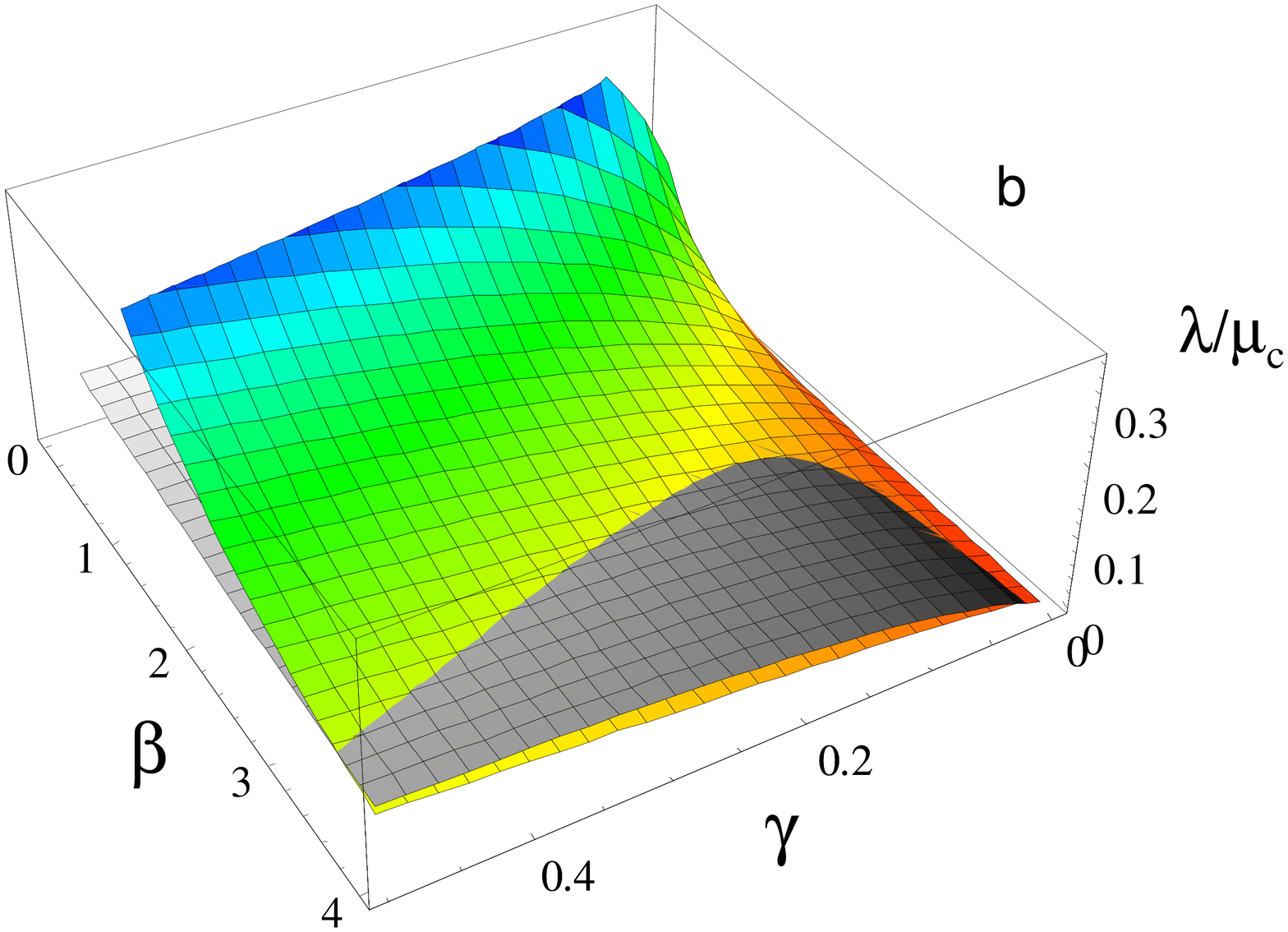}
\caption{The critical values of the electron-phonon coupling,
$\lambda$, for $s$-wave (gray color) and $d$-wave (a) and $p$-wave
(b)  Cooper pairs (bright colors), as  functions of the parameters
$\beta$ and $\gamma$. The Tolmachev-Morel-Anderson logarithm is set
here at $\mu_c\ln(\omega_p/\omega_D)=3$, and $\alpha=3$.}
\end{center}
\end{figure}

 The solution for the $m$-component of the order
parameter ($m=0,\pm 1,\pm 2,...$) is found in the form $ \Delta_{m}
=\Delta^{(1)} _{m}\Theta (\omega _{D}-|\xi_{\bf
k}|)+\Delta^{(2)}_{m}\Theta (\omega _{p}-|\xi_{\bf k} |)\Theta
(|\xi_{\bf k}|-\omega _{D})$ with  different values of $\Delta
^{(1)}_m$ and $\Delta ^{(2)}_m$ below and above the cut-off energy
$\omega _{D}$, respectively. Integrating in Eq.(\ref{master}) over
$\xi _{\bf k^{\prime}}$, $\phi^{\prime}$ and $q_{\perp}$ yields  the
following pair of equations,
\begin{equation} \label{first}
\Delta^{(1)}_m
\left[1-(\lambda_m-\mu_{m})\ln{1.14\omega_D\over{k_BT_c}}\right]
+\Delta^{(2)}_m \mu_{m} \ln{\omega_p\over{\omega_D}}=0
\end{equation}
\begin{equation} \label{second}
\Delta^{(2)}_m \left[1+\mu_{m}
\ln{\omega_p\over{\omega_D}}\right]+\Delta^{(1)}_m
\mu_{m}\ln{1.14\omega_D\over{k_BT_c}}=0.
\end{equation}
Here $\lambda_m$ and $\mu_{m}$ are the phonon-mediated attraction
and the Coulomb pseudopotential in the $m$-pairing channel, given
respectively by
\begin{equation} \label{lambda}
{\lambda_m\over{\lambda}}= \delta_{m,0}+{
\alpha\over{2\sqrt{\gamma}}}
\int_0^{\gamma}{dx[x+1-\sqrt{x(x+2)}]^m\over{\sqrt{x+2}}} ,
\end{equation}
and
\begin{equation} \label{mu}
{\mu_{m}\over{\mu_c}}= {\sqrt{\tilde{\gamma}}\over{2}}
\int_0^{\tilde{\gamma}}{dx[x+\beta+1-\sqrt{(x+\beta)(x+\beta+2)}]^m\over{\sqrt{x(x+\beta)(x+\beta+2)}}},
\end{equation}
where $\lambda= N(0)C^2/NMc_{\parallel}^2$,
$\gamma=\pi^2/2d^2k_F^2(1+\alpha)$,
$\tilde{\gamma}=\gamma(1+\alpha)$, $\mu_c=4e^2d^2N(0)/\pi
V\epsilon_0$, and $\beta=q_s^2/2k_F^2$ (note that $\lambda$,
$\mu_c$, and $q_s$ do not depend on the carrier density since $N(0)$
is roughly constant in the quasi-two dimensional Fermi gas).

The effective attraction of two electrons in the   Cooper pair with
non-zero orbital momentum turns out finite at any finite anisotropy,
$\alpha \neq 0$, but numerically smaller than in the $s$-channel,
Fig.1 (inset) , as  is also seen from its analytical expressions for
$s$-wave pairing, $m=0$ ($\lambda_s$), $p$-wave pairing, $m=1$
($\lambda_p$), and for $d$-wave pairing, $m=2$ ($\lambda_d$),
obtained by integrating in Eq.(\ref{lambda}). When the inter-plane
distance is much larger than the wavelength of electrons, $\gamma
\ll 1$, one obtains $\lambda_s \approx\lambda$, $\lambda_p\approx
\lambda \alpha (\gamma/2)^{1/2}/2$, and $\lambda_d\approx
\lambda_p$. In the opposite limit, $\gamma\gg 1$ one finds
$\lambda_s \approx\lambda (1+\alpha)$, $\lambda_p\approx \lambda
\alpha (2/\gamma)^{1/2}/3$, and $\lambda_d\approx \lambda_p/5$.

The Coulomb repulsion turns out much smaller in the unconventional
pairing states than in the conventional $s$-wave state, Fig.1, which
is also seen from the analytical expression for $\mu_m$,
Eq.(\ref{mu}).  If  $\tilde{\gamma}\lesssim \beta$, the  repulsion
constant $\mu_m$ drops as $1/\beta^{m+1}$ in the $m$-channel at
strong screening, when $\beta \gg 1$. It provides a wide region with
unconventional pairs in the "$\gamma$ - $\beta$" parameter space, in
spite of the lower values of their electron-phonon coupling
constants, Fig.1. Indeed, the critical temperature for $m$-Cooper
pairing is
\begin{equation}
T_{cm}=1.14\omega_D\exp
\left(-{1\over{\lambda_m-\mu_m^\ast}}\right),
\end{equation}
where $\mu _{m}^{\ast }=\mu _{m}/[1+\mu _{m}\ln (\omega _{p}/\omega
_{D})]$, as found from Eq.(\ref{first}) and Eq.(\ref{second}). The
$m$-pairing state with the lowest value of $\lambda/\mu_c$ is the
ground state of the BCS superconductor, where the critical ratio
$\lambda/\mu_c$ is determined by the condition
$\lambda_m=\mu_m^\ast$ as the function of the parameters $\gamma$
and $\beta$.  The critical $d$-wave surface,
$\lambda/\mu_c=S(\gamma, \beta)$, defined by this condition,   is
found below $s$-wave and $p$-wave surfaces, if $\beta\gtrsim 1$, so
that the $d$-wave state is the ground state in this region,  as seen
from Fig.2. Higher momentum states, $m\geqslant 3$, have even a
smaller Coulomb repulsion  at a large $\beta$, Eq.(\ref{mu}), so
that they can be the ground state as well, if $\gamma$ is so small,
that $\lambda_m$ in Eq.(\ref{lambda}) is almost $m$-independent for
$m\geqslant1$. However, an  \emph{in-plane} anisotropy of the sound
velocity, compatible with the symmetry of the perovskite lattice,
could readily stabilise the $d$-wave state against  higher-momentum
states. Naturally, if  the sound speed is enhanced along the
diagonal directions of the in-plane primitive cell, the $d$-wave
order parameter would be zero along  diagonals of the 2D Brillouin
zone, where it changes its sign.

\begin{figure}
\begin{center}
\includegraphics[angle=-90,width=0.50\textwidth]{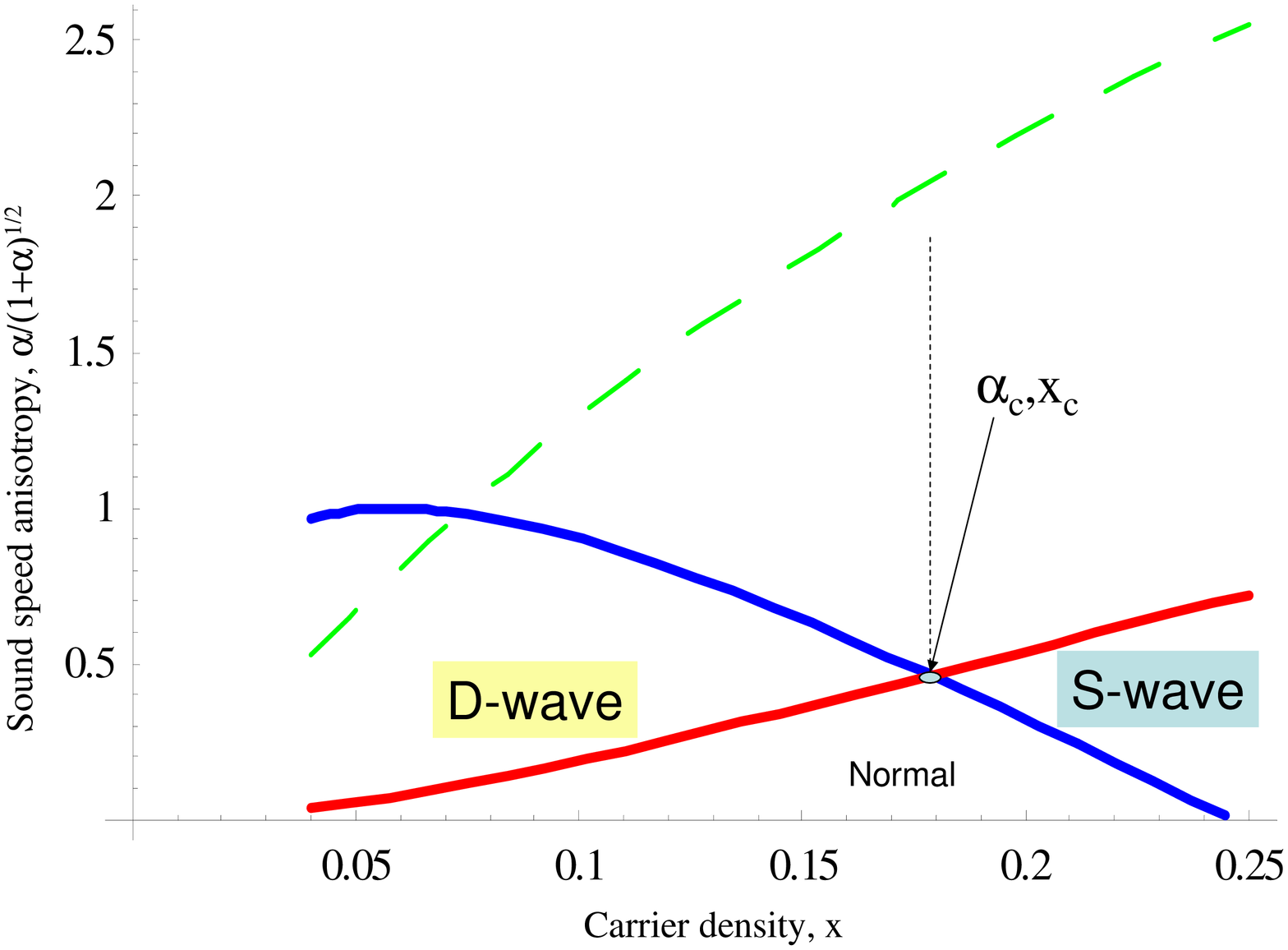}
\vskip -0.5mm \caption{Critical sound-speed anisotropy,
$\alpha/(1+\alpha)^{1/2}=(c_{\parallel}^2-c_{\perp}^2)/c_{\parallel}c_{\perp}$,
as a function of doping, $x$, for $\lambda=\mu_c/12$ (solid lines
correspond to $d$ and $s$ states, and dashed line to $p$-state).
With increasing carrier density there is a quantum phase transition
at $x=x_c$ from a d-wave to an s-wave superconductor, when $\alpha
>\alpha_c$, and  two quantum phase transitions from $d$-wave to
the normal state and from the normal state to the $s$-wave state
when $\alpha < \alpha_c$. }
\end{center}
\end{figure}

Using the simplest parabolic approximation for a 2D-electron energy
spectrum we can draw some conclusions on the carrier-density
evolution of the order-parameter symmetry. Within this
approximation, $k_F^2=2\pi d n$ and $N(0)= mV/2\pi d \hbar^2$, where
$n=2x/\Omega$ is the carrier density and $x$ is the doping level as
in La$_{2-x}$Sr$_x$CuO$_4$ with the unit cell volume $\Omega$. The
ratio of the parameters $\beta=me^2\Omega/2\pi \hbar^2 d^2
\epsilon_0 x$ and $\tilde{\gamma}=\pi\Omega/8 d^3x\thickapprox
0.044/x$   is independent of the carrier density,
$\beta/\tilde{\gamma}=4me^2d/\pi^2\hbar^2\epsilon_0$, which is
approximately $5$ for the  values of $m=4m_e$ and $\epsilon_0=10$.
Fixing the value of the EPI  constant at $\lambda=\mu_c/12$ and
taking  $\mu_c\ln(\omega_p/\omega_D)=3$,  we draw the
\emph{anisotropy-doping } phase diagram, Fig.3, with the critical
lines for $s$, $p$ and $d$ order parameters, defined by
$\lambda_m=\mu_m^\ast$. The state with the lowest magnitude of the
anisotropy, $\alpha/(1+\alpha)^{1/2}$, is the ground state of the
BCS superconductor. At substantial doping the screening length
becomes larger than the typical wavelength of electrons,
$\beta\rightarrow 0$, so that the $s$-wave state is the ground state
at a large number of carriers per unit cell for any anisotropy. On
the contrary, the Coulomb repulsion is reduced to the local
interaction at a low doping, $ \beta\rightarrow \infty$,  and
$d$-wave Cooper pairs are the ground state even at very low value of
the anisotropy, Fig.3. Interestingly, $s-$ and $d$-states turn out
degenerate at some intermediate value of doping, $x=x_c$. Hence
there is a quantum phase transition with increasing doping from $d-$
to $s$-superconducting  state, if $\alpha
> \alpha_c$, and from $d-$ to the normal state and then to the $s$-wave
superconductor, if  $\alpha < \alpha_c$, see Fig.3.

 In the  strong-coupling regime, $\lambda \geqslant 1$, the pairing is
individual \cite{alebook},   in contrast with the collective
 Cooper pairs. While the Bose-condensate of individual bipolarons could break
the symmetry  on a discreet lattice \cite{alesym,andsym}, here I
propose a symmetry breaking mechanism, which works even  in a
continuum model, where the ground state, it would seem, be s-wave
\cite{bristol} to satisfy the theorem \cite{landau}.

The unscreened
 Fr\"ohlich EPI with optical phonons in layered ionic lattices like cuprates
  significantly reduces the finite-range Coulomb
repulsion, since $\epsilon_0 \gg 1$ , allowing the deformation
potential, Eq.(\ref{ph}), to bind  carriers into real-space
bipolarons, if $\lambda \geqslant 1$. The potential, $V({\bf r})=
-\sum_{\bf q} V_{ph}({\bf q})\exp({i{\bf q \cdot r}})$ is non-local
in  real space,
\begin{equation} V({\bf r})= -V_{ph}\Omega
\left[{\delta({\bf r})\over{d}}+{\alpha\over{4\pi
(1+\alpha)^{1/2}r^3}}\right],
\end{equation}
 falling as $1/r^3$ at
the distance $r \gg d$ between two carriers in the plane, where
$V_{ph}=C^2/Mc_l^2$. While its
 local part ($\propto \delta({\bf r})=\delta(x)\delta(y))$ is negated by  the
strong on-site repulsion $U$, the second non-local part provides
bound pairs of different symmetries with the binding energies
$\Delta_s > \Delta_p > \Delta_d >...$ in agreement with the theorem.
However, there is the residual Coulomb repulsion between bipolarons,
$v_c(R)$, significantly reduced by optical phonons. If we
approximate the bipolaron as a point charge $2e$, then
$v_c(R)\approx 4e^2/\epsilon_0 R$. However since bipolarons have a
finite extension, $\xi$, there are corrections to the Coulomb law.
The bipolaron has no dipole moment, hence  the most important
correction at large distances between two bipolarons, $R\gg\xi$,
comes from the charge-quadrupole interaction  \cite{landau},
$v_c(R)=4e^2(1\pm \eta\xi^2/R^2)/\epsilon_0 R$,  where $\eta$ is a
number of the order of 1, and plus/minus signs  correspond to
bipolarons in the same or different planes, respectively. The
dielectric screening, $\epsilon_0$ is highly anisotropic in
cuprates, where the in-plane dielectric constant,
$\epsilon_{0\parallel}$, is much larger then the out-of-plane one,
$\epsilon_{0\perp}$ \cite{dielectric}. Hence the inter-plane
repulsion provides the major contribution to the condensation
energy. Since $\xi^2 \propto 1/\Delta$, the repulsion of
unconventional bipolarons with smaller binding energies, $\Delta_d,
\Delta_p < \Delta_s$, is reduced compared with the repulsion of
$s$-wave bipolarons. As a result, with increasing carrier density we
anticipate a transition from  BEC of $s$-wave bipolarons to BEC of
more extended $p-$ and $d$-wave real-space pairs in the
strong-coupling limit.

A great number of observations, in particular  phase-sensitive
experiments \cite{pha}, point to the unconventional $d$-wave
symmetry of  cuprate superconductors (for a review see \cite{ann}).
Several authors \cite{band} have remarked that superexchange, and
not phonons is responsible for this symmetry breaking. Here I arrive
at the opposite conclusion. Indeed, superexchange interaction, $J$,
is proportional to the electron hopping integral, $t$,  squared
divided by the on-site Coulomb repulsion (Hubbard $U$), $J=4t^2/U$,
estimated as $J \thickapprox 0.15$ eV in  cuprates \cite{band}. This
should be compared with the acoustic-phonon pairing interaction,
$V_{ph}$, which is roughly the Fermi energy, $V_{ph}\thickapprox
E_F\thickapprox 4t$ in a metal \cite{bardeen}, or the bandwidth
squared divided by the $ion-ion$ interaction energy of the order of
the \emph{nearest-neighbour} Coulomb repulsion, $Mc_l^2 \thickapprox
V_c$ in a doped insulator \cite{anselm}. The small ratio of two
interactions, $J/V_{ph}\thickapprox t/U \ll 1$, or
$J/V_{ph}\thickapprox V_c/U \ll 1$ and the giant sound-speed
anisotropy \cite{migliori,chang} favor conventional EPI  as the only
origin of the unconventional pairing both  in underdoped cuprates,
where the pairing is individual \cite{alebook}, and in overdoped
samples apparently with  Cooper pairs \cite{bcs}.

I thank  A. F. Andreev, I. Bozovic, J. P. Hague,   V. V. Kabanov, P.
E. Kornilovitch, K. I. Kugel, M. L. Kulic, I. I. Mazin, J. H.
Samson, and L. M. Satarov for valuable discussions. The work was
supported by EPSRC (UK) (grant Nos. EP/C518365 and EP/D035589).

\end{document}